\documentclass{iau} 
\usepackage{graphicx}

\title[IAU Symp. 353.~~ETGs/Stellar halos connection] 
{Elliptical galaxies/stellar halos connection}

\author[Magda Arnaboldi, Claudia Pulsoni, Ortwin Gerhard et al. ]   
{Magda Arnaboldi$^1$, Claudia Pulsoni$^{2,3}$ , Ortwin Gerhard$^2$\\
 \and the ePN.S team}

\affiliation{$^1$ European Southern Observatory, K. Schwarzschild
  Str. 2, DE-85748, Garching, Germany
  \\ email: {\tt marnabol@eso.org} \\[\affilskip]
  $^2$ Max-Planck-Institut f\"ur extraterrestrische Physik, Giessenbachstrasse, DE-85748 Garching, Germany; email: {\tt cpulsoni@mpe.mpg.de; gerhard@mpe.mpg.de}\\[\affilskip]
  $^3$Excellence Cluster Universe, Boltzmannstrasse 2, DE-85748,
  Garching, Germany }

\pubyear{2019}
\volume{353}  
\jname{Galactic Dynamics in the Era of Large Surveys}
\editors{M. Valluri, \& J. A. Sellwood  }
\begin{document}

\maketitle

\begin{abstract}Cosmological simulations predict that early-type
  galaxies (ETGs) are the results of extended mass accretion
  histories. The latter are characterized by different numbers of
  mergers, mergers’ mass ratios and gas fractions, and
  timing. Depending on the sequence and nature of these mergers that
  follow the first phase of the {\it in-situ} star formation, these accretion
  histories may lead to ETGs that have low or high mass halos, and
  that rotate fast or slow. Since the stellar halos maintain the
  fossil records of the events that led to their formation, a
  discontinuity may be in place between the inner regions of ETGs and
  their outer halos, because the time required for the halos’ stars to
  exchange their energies and momenta is very long compared with the
  age of these systems. Exquisite deep photometry and extended
  spectroscopy for significant samples of ETGs are then used to
  quantify the occurrence and significance of such a transition in the
  galaxies’ structural and kinematical parameters. Once this
  transition radius is measured, its dependency with the effective
  radius of the galaxies’ light distribution and total stellar masses
  can be investigated. Such correlations can then be compared with the
  predictions of accreted, i.e. {\it ex-situ} vs. {\it in-situ}
  components from cosmological simulations to validate such models.
\end{abstract}

\firstsection 
\section{Introduction}
Our most recent understanding of the physics and structure of the
brightest regions of elliptical galaxies, hereafter early-type
galaxies or ETGs, comes from integral field spectroscopic data
collected in magnitude limited surveys of galaxies, in the nearby
universe.  From the ATLAS3D survey (\cite[Cappellari et
al. 2011]{Cappellari+2011}) to the more recent results from the MANGA
survey (\cite[Graham et al. 2018]{Graham+2018}) a paradigm emerged
that groups ETGs in either fast (FRs) and slow rotators (SRs),
according to the central projected specific angular momentum within
one effective radius (\cite[Emsellem et al. 2007]{Emsellem+07}). FRs
do include S0 galaxies; the FR class encompasses the great majority
of the ETGs (86\%). Their brightest regions appear as oblate systems
with regular disk-like kinematics along the photometric major
axis. Furthermore their formation history is dominated by gas-rich
processes. SRs, on the other hand, are $\simeq 16\%$ of the ETGs,
often display kinematic features such as counter-rotating cores or
twists of the kinematic position angle. These objects are relatively
rounder systems, mildly triaxial (\cite[Foster et
al. 2017]{Foster+17}), tend to be massive (\cite[Cappellari et al.
2013]{Cappellari+2013}), and their formation history is dominated by
dry mergers.

Observations of ETG analogs at redshifts $z=2-3$ show significant
structural difference though from their local counterparts. In this
redshift range, massive $\sim 10^{11} M_\odot$ systems are flatter,
have effective radii less than $2$\,kpc and their surface brightness
profiles are modeled with a Sersic profile whose index $n$ values are
lower than those of local ETGs (\cite[Van Dokkum et
al. 2010]{vDokk+10}, \cite[Toft et al. 2007]{Toft+07}). In turn, local
(z=0) massive galaxies have grown by a factor two in mass, four in
size, with their effective radii
$R_e \propto M^\alpha (\alpha \ge 2)$. In addition they have surface
brightness profiles whose Sersic index $n$ value is $>5$ and with rounder
isophotes (\cite[Kormendy et al. 2009]{Kormendy+2009}).

Simulations explain such growth via dry (gas poor) minor mergers onto
an existing earlier formed host within its dark halo (\cite[Oser et
al. 2010]{Oser+10}, \cite[Hilz et al. 2013]{Hilz+13}), also refereed
to as ``two-phase formation scenario''. Depending on the mass
ratio/merging parameters, these mergers leave the inner host structure
almost unchanged (\cite[Amorisco 2017]{Amorisco17}). In case of
mini-mergers ($M_{sat}:M_{host} > 1:10$) the stars stay at large radii
and never come near to the center of the potential (\cite[Karademir et
al. 2019]{kara+19}). These events increase size growth and cause
faster shape change compared to mergers with higher mass satellites.

\begin{figure}[b]
\begin{center}
\includegraphics[width=3.4in]{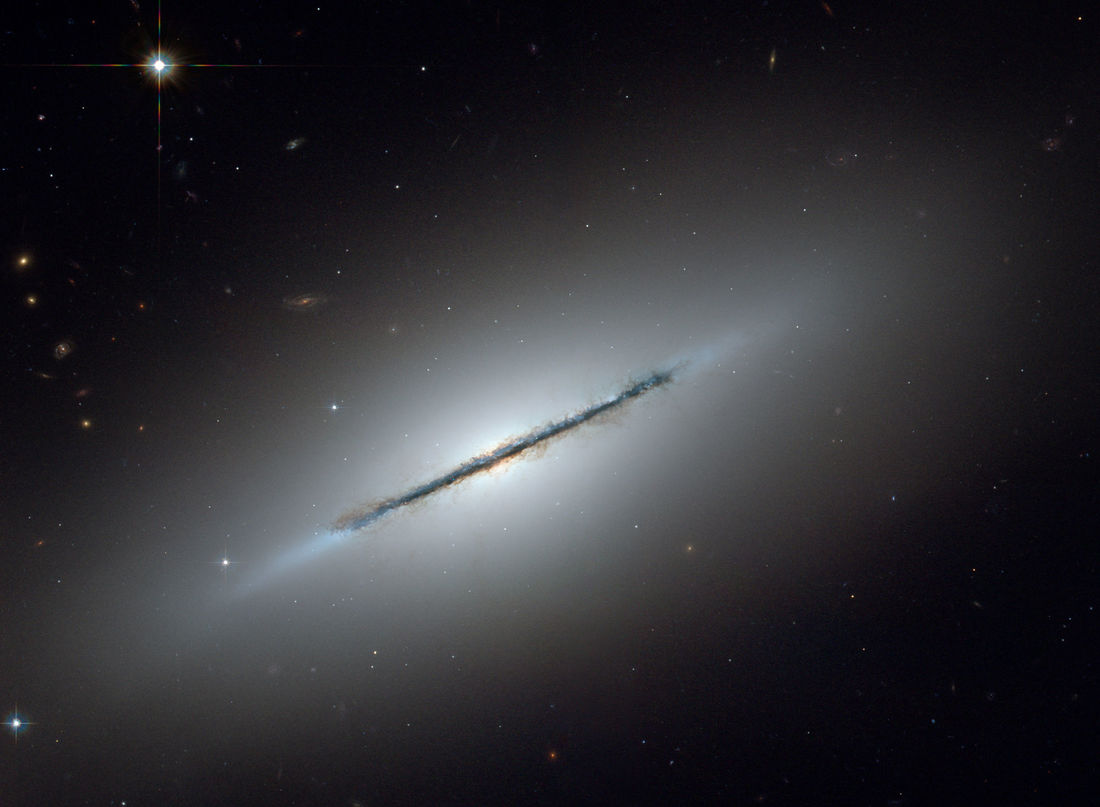} 
\includegraphics[width=3.4in]{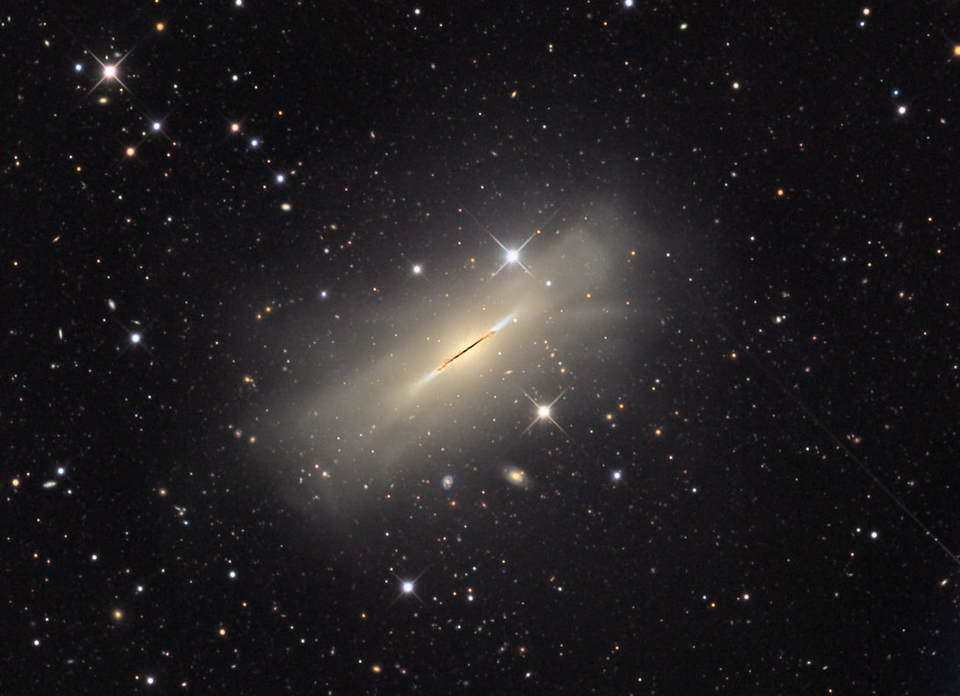} 

\caption{Top: The nearby galaxy NGC 5866 is a disk galaxy seen almost
  exactly edge-on. Credit: NASA, ESA, and The Hubble Heritage Team
  (STScI/AURA). FoV is $\sim 3' \times 3'$. Bottom: A deep image of
  NGC 5866 reveals huge streams of stars looping around it; clear
  evidence of recent accretion events. Credit: Adam Block/Mount Lemmon
  SkyCenter/University of Arizona. FoV is $\sim 35' \times 35'$, North
  is up and East to the left. }
   \label{fig1}
  \end{center}
\end{figure}

One can reach an intuitive understanding of these accretion processes
by looking at the $z=0$ analogs of such dry merger events. A ``poster
child'' is the spindle galaxy NGC~5866. This galaxy is at 13.4 Mpc
distance and has a $R_{25} = 10$\,kpc. The morphology of the high
surface brightness regions of this galaxy (see Fig.\,\ref{fig1}) is
SA0 with a narrow dust lane seen edge-on and forming an angle with the
photometric major axis. In the deep HST image, the dust filaments
appear orthogonal to the disk. The kinematics of this galaxy is that
of a regular disk rotator (\cite[Krajnovi´c \etal\ 2011]{Krayj+11}). The
deep image of NGC~5866 shows that something different is happening in
its outskirts. In Fig.\,\ref{fig1} at low surface brightness the
galaxy has extra light in the halo from a dissolving satellite on
an extended nearly-radial orbit, with several pericenter passages. Clearly
when such an accretion event is phase-mixed, it will add to this
galaxy a different shape component from the central one with the FR
kinematics.

ETGs like NGC~5866 are ``cosmological'' in the sense that they have
accreted components that form their halos. Thus the variety and scatter in
the physical properties of these accretion events and their final
results, i.e. the halos in ETGs, are tied to their cosmological
context. This is reflected in the halo kinematics which is the subject
covered by this work.

\section{Probing ETG halo kinematics with Planetary Nebulae}

In the cosmological context of the formation of ETGs, it is
important to measure the kinematics of their stellar halos because
{\it i) } they contain $>50\%$ of the stellar angular momentum, and
{\it ii)} $95\% $ of their total mass, mostly dominated by dark
matter. Since {\it iii)} the stellar content of the halos are mostly
accreted (\cite[Wu et al. 2014]{Wu+14}) and {\it iv)} the settling
time scale are long ($\sim 1$ Gyr; \cite[Bullock \& Johnston,
2005]{B+J05}) their kinematics preserve the signatures of the
formation processes. See for example the effects of the satellite
accretion, i.e. the Magellanic Clouds, on the dark halo potential of
the host, i.e. the MW (Contribution by G. Besla, this conference).

These goals are complementary to those of the completed (ATLAS3D) or
on-going IFS surveys (MANGA, CALIFA, SAMI, MASSIVE) that are
targeting the central $1-2\, R_e$. To measure the kinematics of the
stellar halos at large distances we use Planetary Nebulae (PNe) in
these halos. Planetary nebulae are relatively bright [OIII] 5007 \AA\
emitters that become easily detectable in the dimmed low surface
brightness halo regions. PNe are the late phases of intermediate to
low mass stars and provide discrete sampling of the line-of-sight
velocity distribution (LOSVD) in any region of a given nearby galaxy
(\cite[Arnaboldi et al. 2002]{Arna+02}). In all galaxies studied so
far, whenever PNe and stellar kinematics from absorption line
spectroscopy are available, their agree. PNe are therefore good
kinematic tracers.

\subsection{The extended PN.S ETG survey}
The extended Planetary Nebulae Spectrograph (PN.S) ETG survey
(\cite[Arnaboldi et al. 2017]{Arna+17}) targeted 33 ETGs in a
magnitude limited sample with a wide range of structural parameters
(luminosity central velocity dispersion ellipticity surface
brightness). Observations were carried out mostly with the PN.S
mounted on WHT at the ING observatory on La Palma. The PN.S is a
custom-built slitless spectrograph optimized for the detection of the
monochromatic [OIII] emission from PNe at 5007 \AA\ (\cite[Douglas et
al. 2007]{ND+07}).  With the PN.S, the detection of a PN [OIII] 5007
\AA\ emission and its LOSV measurement can be achieved in a single
observing run. In the ePN.S survey, LOSVs were measured for 8636
extragalactic PNe, the largest sample available so far, with samples
of $80-700$ PN LOSVs per galaxy. The ETGs in the ePN.S sample are
within 25 Mpc distance, with 24 FRs and 9 SRs. The kinematics mapped
by PNe covers the radial range out to $3-13 R_e$ with $5.6 R_e$ as
median value for the entire sample (\cite[Coccato et al 2009]{Coc+09},
\cite[Cortesi et al. 2013]{Cort+13}; \cite[Pulsoni et
al. 2018]{Puls+18}). For the complete description of the smoothing
technique for the extraction of the kinematics maps from discrete PN
LOSV we refer to \cite[Pulsoni et al. 2018]{Puls+18}.

\section{Kinematical diversity of stellar halos}

\begin{figure}
\begin{center}
\includegraphics[width=3.4in]{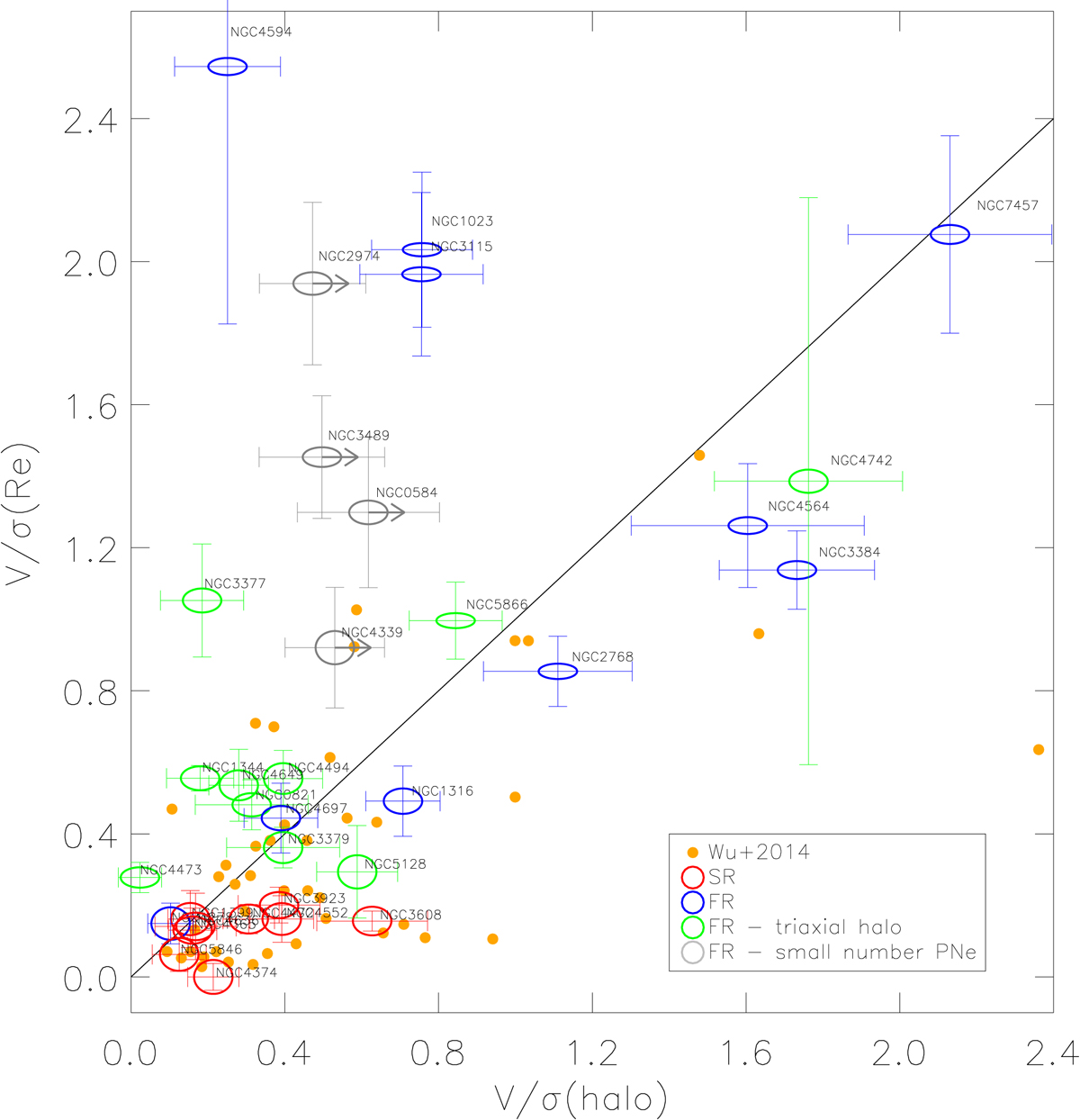} 

  \caption{$V/\sigma(R_e)$ from absorption line data compared with
    $V/\sigma(halo)$ from PN data; from \cite[Pulsoni et
    al. 2018]{Puls+18}.  ePN.S FRs and SRs are shown with different
    colors, as are the FR with triaxial halos. The gray open ellipses
    represent the galaxies (all FRs) with fewer tracers, for which the
    ePN.S analysis provides a lower limit to the $V/\sigma(halo)$
    value. The solid line traces equal values of $V/\sigma(R_e)$ and
    $V/\sigma(halo)$. The orange circles shows the $V/\sigma$ values
    of the simulated galaxies from \cite[Wu et al. (2014)]{Wu+14},
    re-scaled by the appropriate quantity (see \cite[Pulsoni et
    al. 2018]{Puls+18} for further details) . While the SRs show
    increased rotation in their halos, the majority of the FRs show a
    drop in rotation at large radii. }
   \label{fig2}
  \end{center}
\end{figure}

The results from the ePN.S survey show that the kinematics of ETGs at
large radii can be substantially different from that of the inner
regions. For the SRs, the rotation increases in the halos. For the
FRs, this diversity is signaled by a decrease in rotation in 70\% of
the ePN.S FRs or by a significant twist of the kinematical position
angle in 40\% of the ePN.S FRs. 30\% of the ePN.S FRs are rapidly
rotating at large radii also. We interpreted the general behavior of
the ePN.S FRs as the transition from the inner disk component to the
spheroidal halo, which is dispersion dominated and might deviate from
axisymmetry.  Figure~\ref{fig2} illustrates this inference by
comparing the $V/\sigma$ of the halo with that of the inner regions at
$1\,R_e$. The $V/\sigma(R_e)$ values are derived by interpolating the
$V_{rot}$ and $\sigma$ profiles from integrated light at $R = R_e$,
while $V/\sigma(halo)$ is the ratio of $V_{rot}$ and $\sigma$
estimated in the outermost radial bin of the PN velocity fields.  All
the SRs are located below the $1:1$ line, showing higher rotational
support at large radii. The scatter of the FRs in this diagram
reflects their different intrinsic structure and kinematics.  The
halos of most ePN.S FRs have $V/\sigma$ ratio similar to that of
SRs. Among these sub-sample of FRs the scatter in $V/\sigma(R_e)$ is
probably driven by the presence of a more or less prominent disk
component seen at different inclinations, and embedded into a 
dispersion dominated spheroid. The flattening of the ellipses in
Fig.~\ref{fig2} indeed shows that galaxies with higher $\epsilon$
also display higher $V/\sigma(R_e)$.  A second group of FRs with high
rotational support in the halo populate the diagram on the right of
the $1:1$ line.  These galaxies are either dominated by disk rotation at
all radii (NGC~3384 and NGC~7457), or have a rapidly rotating spheroid
(NGC~2768), or either of these (NGC~4564, NGC~4742).  The FRs with
triaxial halos typically show equal values of $V/\sigma(R_e)$ and
$V/\sigma(halo)$, spanning all values in $V/\sigma(halo)$.

\section{Kinematic Transition Radius}

We found that most of the galaxies in the ePN.S sample show a
kinematic transition signaled by a variation of the rotation velocity
$V_{rot}$ or in the kinematic position angle. In the framework of the
two-phase formation scenario, this kinematic difference can be
understood in association with stellar components of different
origins, such as the {\it in-} and the {\it ex-situ} components.
Predictions from simulations (\cite[Rodriguez-Gomez et al
2016]{RG+2016}) prescribe the {\it in-situ} stars to be concentrated
in the central regions of galaxies, while the accreted stars dominate
the halos, and with their relative contribution gauged by the the
total mass. Within this framework we can define a
\underline{transition radius} $R_T$ as the distance where we measure a
quantifiable variation of the stellar kinematics, and then compare it
with the total stellar mass.  The transition radius and its
uncertainty $R_T\pm \Delta R_T $ are measured using (A) the interval
between the radius at which $V_{rot}$ is maximum and the radius at
which it decreases by $\sim 50$\,kms$^{-1}$; (B) in case of an
increasing $V_{rot}$ profile, the radial range in which $V_{rot}$
increases from $\sim 0$ to $\sim 50$\,kms$^{-1}$; (C) in case of a
kinematic twist, the radial range in which the position angle changes
significantly.

\begin{figure}
\vspace*{-0.5 cm}
\begin{center}
\includegraphics[width=3.4in]{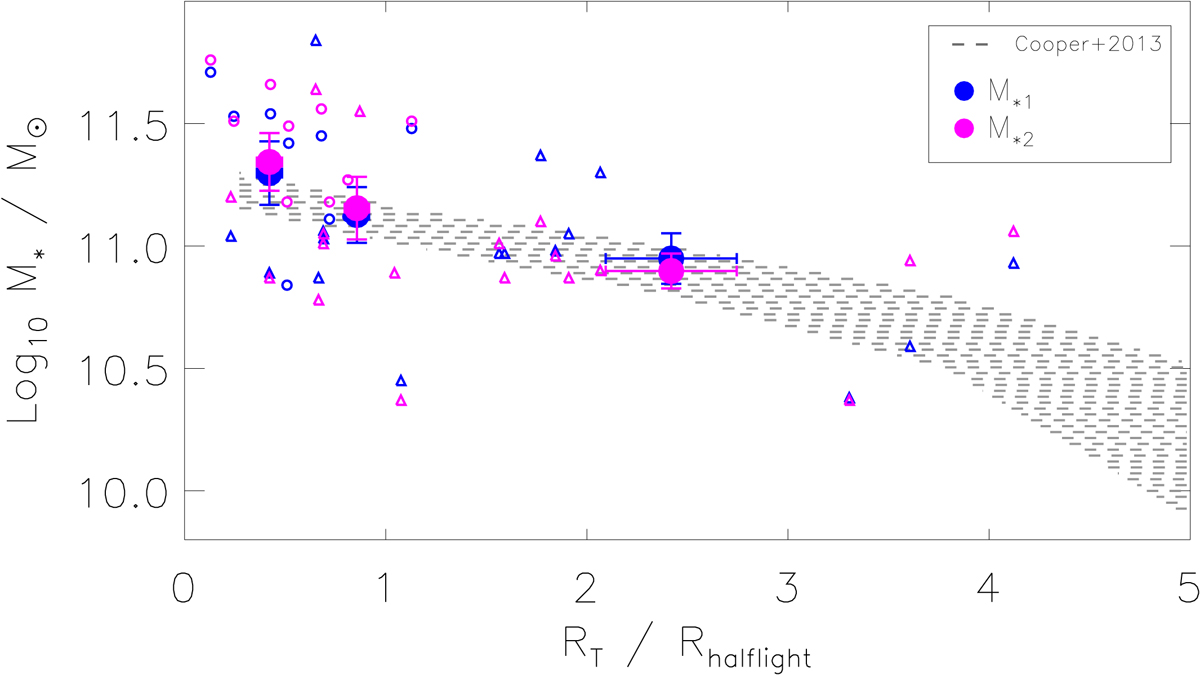} 
  \caption{Transition radius in units of $R_{halflight}$ versus total
    stellar mass (open symbols: circles for SRs, triangles for
    FRs). The full symbols show the same quantities in bins of
    $R_T=R_{halflight}$. Different colors show the results of two
    different procedures for calculating the total stellar mass. From
    \cite[Pulsoni et al. 2018]{Puls+18}. }
   \label{fig3}
  \end{center}
\end{figure}

Figure~\ref{fig3} shows the stellar mass of the galaxies versus
$R_T/R_{halflight}$, where $R_{halflight}$ is the half-light radius
measured from the light growth curves. The full circles show the same
quantities in three bins of $R_T/R_{halflight}$; the error bars
represent the standard deviation of the mass and of the
$R_T/R_{halflight}$ ratio in the bin. For the tabulated values we
refer to \cite[Pulsoni et al. 2018]{Puls+18}. A clear correlation
exists between total stellar mass and $R_T/R_{halflight}$, in the
sense that the more massive galaxies tend to have transition radii at
smaller fractions of $R_{halflight}$. The shaded region in
Fig.\,~\ref{fig3} shows the corresponding quantities from the N-body
simulations of \cite[Cooper et al. (2013)]{Coop+2013}. Fig.~\ref{fig3}
illustrates an overall agreement of the predictions from cosmological
simulations with the structures of the ETG halos as constrained by PN
kinematics. 

From cosmological simulations, we also learn that extended accretion
of stars in massive ETGs occurs preferentially on radial orbits
(\cite[R\"ottgers et al. 2014]{Roett+14} and reference therein) which
may lead to a change of shape toward a more round-mildly triaxial shape
at larger radii. The imprints of radial orbits on the higher order
moments of the LOSVDs were modeled for M49 and M87 at $70-100$ kpc
distance using PNe (\cite[Longobardi et al. 2018]{Long+18},
\cite[Hartke et al. 2018]{Hartke+18}).

\section{Conclusions}
The ePN.S is the largest survey of extragalactic PNe. PNe are evolved
low mass stars and reliable tracers of the kinematic of their parent
stellar populations, hence they can be used to probe the outer regions
of the halos. We measured the kinematics in ETGs out to $\sim 6 R_e$
in 33 ETGs.

ETGs have more diverse kinematic properties than in the central
regions. For the ePN.S sample we find 1) the onset of rotation in the
halos for the SRs. 2) For the FR, 70\% have slowly rotating halo,
while in 30\% of the sample the halo rotate as fast as the inner
regions. 3) In 40\% of the ePN.S ETGs the outer halo is triaxial
(consistent with the photometric twist of the isophotes at large
radii).

The kinematic transition radius anti-correlates with stellar mass, in
agreement with predictions with cosmological particle tagging
simulations.

The ePN.S FRs with triaxial halo are among the most massive galaxies
for which the accreted component is expected to be more prominent.

Outer stellar halos are indeed linking galaxies with their
cosmological environment.

\section{Acknowledgments} 
MAR wished to thank the SOC of the IAU Symp. 353 for the invitation to
give this talk. The authors are grateful to Nigel G. Douglas for his
fundamental contribution to the foundation of the Planetary Nebula
Spectrograph instrument and acknowledge the support and advice
of the ING staff.

\end{document}